# MAGNETISATION REVERSAL AND DOMAIN STRUCTURE IN THIN MAGNETIC FILMS: THEORY AND COMPUTER SIMULATION

U. Nowak

Theoretische Tieftemperaturphysik, Gerhard-Mercator-Universität Duisburg, 47048 Duisburg, Germany

*Abstract*— A model is introduced for the theoretical description of nanoscale magnetic films with high perpendicular anisotropy. In the model the magnetic film is described in terms of single domain magnetic grains, interacting via exchange as well as via dipolar forces. Additionally, the model contains anisotropy energy and a coupling to an external magnetic field. Disorder is taken into account in order to describe realistic domain and domain wall structures. Within this framework the dependence of the energy on the film thickness can be discussed. The influence of a finite temperature as well as the dynamics can be modeled by a Monte Carlo simulation. The results on the hysteresis loops, the domain configurations, and the dynamics during the reversal process are in good agreement with experimental findings.

## I. Introduction

In ferromagnetic films two different mechanisms can be thought of to dominate the reversal process: either nucleation or domain wall motion [1]. Which of these mechanisms dominates a reversal process depends on the interplay of the different interaction forces between domains with different magnetic orientation. In a recent experiment on $Co_{28}Pt_{72}$ alloy films a crossover from magnetisation reversal dominated by domain growth to a reversal dominated by a continous nucleation of domains was found depending on the film thickness which was varied from 100Å to 300Å. Correspondingly, characteristic differences for the hysteresis loops and the dynamics of the reversal process were found [2]. For the case of domain growth the driven domain wall was found to be rough, i.e. self affine.

The goal of this paper is to understand the dependence of the reversal mechanisms on the film thickness and to simulate a corresponding micromagnetic model by Monte Carlo methods [3].

Manuscript received February 13, 1995.
U. Nowak, e-mail uli@thp.uni-duisburg.de, fax +49 203-379-2965;
This work was supported by the Deutsche Forschungsgemeinschaft through Sonderforschungsbereich 166

## II. A Micromagnetic Model

$Co_{28}Pt_{72}$ alloy films have a polycrystalline structure with grain diameters of 100-250Å. For a theoretical description [4] the film is thought to consist of cells on a square lattice with a square base of size $L^2$ where $L = 175$Å. The height $h$ of the cells varies from 100Å to 300Å. Due to the high anisotropy of the $Co_{28}Pt_{72}$ alloy film the grains are thought to be magnetised perpendicular to the film only with a uniform magnetisation $M_s$ which is set to half of the experimental value of $M_s = 360 kA/m$ for the saturation magnetisation in these systems [5] in order to modulate the fact that the magnetisation of a cell is reduced when it is surrounded by a Bloch-like domain wall. The grains interact via domain wall energy and dipole interaction. The coupling of the magnetisation to an external magnetic field $H$ is taken into acount as well as an energy barrier which has to be overcome during the reversal process of a single cell.

From these considerations it follows that the energy needed to reverse a cell $i$ with magnetisation $L^2 h M_s \sigma_i$ with $\sigma_i = \pm 1$ is:

$$\Delta E_i = \sum_{<i,j>} L h S_w \sigma_i \sigma_j - \mu_0 M_s^2 L h^2 \sum_j \frac{\sigma_i \sigma_j}{r_{i,j}^3} + 2\mu_0 H L^2 h M_s \sigma_i + \delta_i \quad (1)$$

The first term describes the wall energy $\Delta E_w$. The sum is over the four next neighbors and $S_w = 0.004 J/m^2$ is the Bloch-wall energy density which follows from the experimental values for the anisotropy constant and the exchange parameter.

In the second term describing the dipole coupling $\Delta E_d$ the sum is over all cells. $r_{i,j}$ is the distance between two cells $i$ and $j$ in units of the lattice constant $L$. The dependency of the dipole interaction on the distance $r_{ij}$ is an approximation for large distancies. For smaller distancies the correct form can be determined numerically and was taken into account.

The third term describes the coupling $\Delta E_H$ to an external field $H$.

The last term is the energy barrier mentioned above. If the reversal mechanism for a single cell is a rotation of the magnetisation vector described by an angel $\theta$, the anisotropy leads to an energy barrier $L^2 h K_u \sin^2(\theta)$.

The other parts of the energy are proportional to $\cos(\theta)$. Therefore $\Delta E_i$ was taken to be the maximum of the function $(\Delta E_w + \Delta E_d + \Delta E_H)\cos(\theta) + K_u \sin^2(\theta)$ with $K_u = 50 kJ/m^3$.

In order to simulate realistic domain structures disorder has to be considered since in real materials the grain sizes are randomly distributed. In the model above this would correspond to a random distribution of $L$ which, however, can hardly be simulated. Therefore, as a simplified ansatz to modulate disorder the variables $\sigma_i$ were randomly distributed around the mean value $|\sigma| = 1$. The distribution is gaussian with width 0.1.

It should be emphasized that all energies in (1) scale linearly with $h$ apart from the dipole energy which scales with $h^2$. This is obviously the reason for the qualitative change in the reversal mechanisms of magnetic films — in thicker films the influence of dipolar forces is stronger.

The simulation of the model above was done via Monte Carlo methods [3], [6]. The flip probability for a randomly choosen cell $i$ was taken to be $\exp(-\Delta E_i)$ for $\Delta E_w + \Delta E_d + \Delta E_H > 0$ and $\exp(-\delta_i)$ for $\Delta E_w + \Delta E_d + \Delta E_H < 0$. This corresponds to a Metropolis algorithm with an additional energy barrier $\delta_i$. The algorithm satisfies detailed balance and Glauber dynamics consequently.

The size of the lattice was typically $150 \times 150$. The dipole interaction was taken into account rigorously without any cut-off or approximation. The boundary conditions were open.

### III. Results

Many of the experimental findings can be described and understood by the model introduced above, especially the dependence of the magnetisation reversal mechanism on the film thickness. This crossover is due to the larger influence of the dipole interaction in the thicker sample. The dipole interaction destabilizes the system leading to an enhanced nucleation rate during the reversal process. In the following, results are shown only for the two cases of a rather thin film (100Å) and a thicker film (300Å) allthough other film thicknesses in between this region have been considered as well.

Fig. 1 shows a simulated domain configuration for the case of a 100Å film. Black regions show reversed cells. The reversal process starts obviously at only a few nuclei from which domains with a more or less circular shape start to grow. The domain wall can be understood as a driven interface in a random medium. The reason for the roughness of the domain wall can be found in the kinetics of the domain wall motion as well as in the disorder of the medium [7], [8].

Fig. 2 shows a domain configuration during the reversal of a thicker film (300Å). The magnetisation is nearly the same as in the picture above but now the reversal is dominated by an increasing number of nuclei. Both do-main configurations are in very good agreement with the pictures from polarization microscopy based on the polar magnetooptic Kerr effect [2].

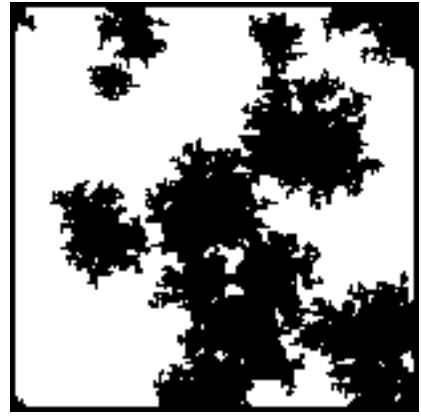

Fig. 1. Simulated domain configuration during the reversal for a 100Å film.

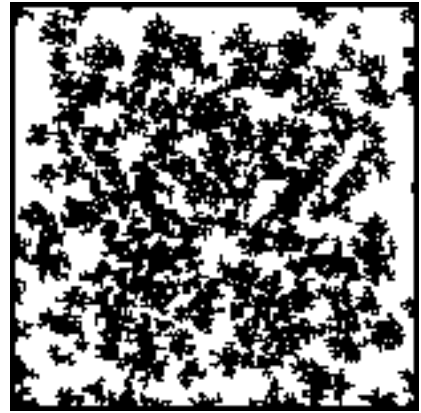

Fig. 2. Simulated domain configuration during the reversal for a 300Å film.

Fig. 3 shows the corresponding simulated hysteresis loops for both magnetic films. For the thin film the hysteresis loop is nearly rectangular as can be explained through the reversal mechanism: once a nucleus begins to grow the domain wall motion does not stop until the magnetisation has completely changed. For the thicker film the enhanced dipolar forces stabilize a mixed phase which can be changed only by a further increase of the external field. The agreement of the shape of the hysteresis loop with experimental results [2] is very good apart from the fact that the critical fields are much to large in the simulation. This point is still under investigation.

The different reversal mechanisms also manifest themselves in a change of the dynamical behavior which is shown in Fig. 4. For the case of nucleation driven reversal there is a rapid change of the magnetisation at the beginning of the reversal process. For the case of do-main growth the change of magnetisation is slower at the beginning since if a constant domain wall velocity is assumed, the area of a single nucleus grows slowly when the nucleus is small, i.e. at the beginning of the reversal process. The time that is needed for a reversal depend on the strength of the driving magnetic field, the temperature and, of course, on the film thickness. For the case shown in Fig. 4 the reversal time is approximately 400MCS for the 100Å film and 15MCS for the 300Å film.

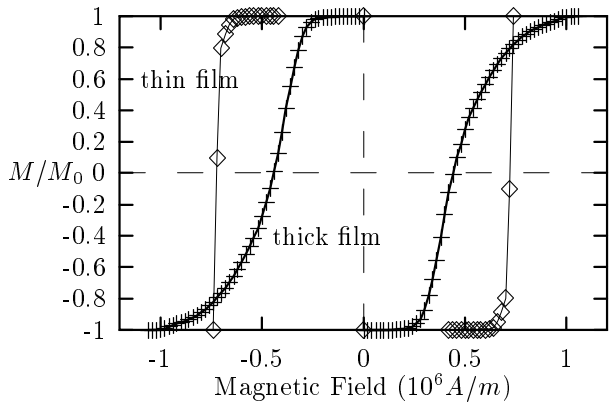

Fig. 3. Magnetisation versus magnetic field during field reversal for the thin and the thick film.

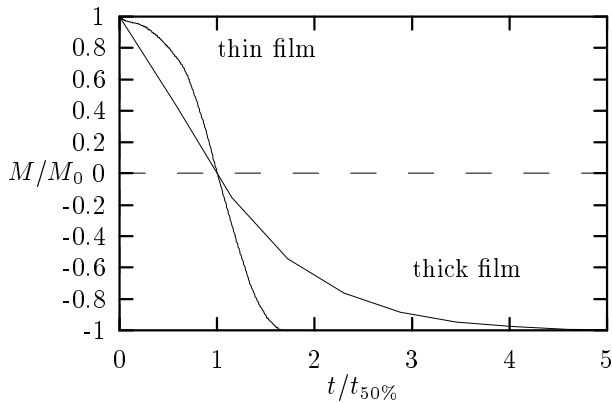

Fig. 4. Magnetisation versus time after a rapid quench to large negative field.

## IV. Conclusions

In this paper the change of the reversal mechanisms in thin magnetic films with perpendicular anisotropy from domain growth to nucleation depending on the film thickness is explained using a rather simple micromagnetic model. It should be emphasized however that this model contains much more information: generally, in the model above the change of reversal mechanisms is attributed to a change in the ratio of the different contributions to the energy in (1). Motivated by recent experiments only the influence of a variation of the film thickness is discussed, allthough a change of other properties of the model like the wall energy $S_w$ or the grain size $L$ would also lead to a change in the ratio of the different energy contributions resulting in effects which can also be described using this model.

Apart from this, additional important properties to be considered are the temperature dependence of the reversal process and the quantitative description of the domain wall, i.e. the roughness or fractal dimension and the dynamics of the wall. These interesting quantities are left for future research.

## Acknowledgment

The author thanks T. Kleinefeld and K. D. Usadel for fruitful discussions.